\documentclass[a4paper,USenglish,cleveref,autoref]{lipics-v2019}

\nolinenumbers

\usepackage{tikz}
\usetikzlibrary{calc,positioning,patterns,snakes}
\definecolor{coqgreen}{rgb}{0.1,0.8,0.0}

\usepackage[frozencache, cachedir=.]{minted}
\setminted{fontsize=\small}
\def\coqin#1{\mintinline{ssr}{#1}}

\makeatletter
\AtBeginEnvironment{minted}{\dontdofcolorbox}
\def\dontdofcolorbox{\renewcommand\fcolorbox[4][]{##4}}
\makeatother

\makeatletter
\let\FV@ListProcessLineOrig\FV@ListProcessLine
\def\FV@ListProcessLine#1{%
  \ifx\FV@Line\empty
    \hbox{}\vspace{-1ex}%
  \else
    \FV@ListProcessLineOrig{#1}%
  \fi}
\makeatother

\BeforeBeginEnvironment{minted}{\smallskip}
\def\coq{\textsc{Coq}}
\def\ssreflect{\textsc{SSReflect}}

\def\rank{\textsf{rank}}
\def\select{\textsf{select}}
\def\delete{\textsf{delete}}
\def\pred{\textsf{pred}}
\def\succ{\textsf{succ}}
\def\vecset{\textsf{set}}
\def\vecclear{\textsf{clear}}
\def\mydef#1{{\sl #1}}
\def\us{\char`\_}

\title{Proving tree algorithms for succinct data structures}

\author{Reynald Affeldt}{National Institute of Advanced Industrial Science and Technology, Japan}{reynald.affeldt@aist.go.jp}{http://orcid.org/0000-0002-2327-953X}{}

\author{Jacques Garrigue}{Graduate School of Mathematics, Nagoya University, Japan}{garrigue@math.nagoya-u.ac.jp}{https://orcid.org/0000-0001-8056-5519}{}

\author{Xuanrui Qi}
{Department of Computer Science, Tufts University, United States}{xqi01@cs.tufts.edu}{https://orcid.org/0000-0002-2032-1552}{}

\author{Kazunari Tanaka}{Graduate School of Mathematics, Nagoya University, Japan}{tanaka.kazunari@k.mbox.nagoya-u.ac.jp}{https://orcid.org/0000-0001-8460-7763}{}

\authorrunning{R.\ Affeldt, J.\ Garrigue, X.\ Qi, and K.\ Tanaka}

\Copyright{Reynald Affeldt, Jacques Garrigue, Xuanrui Qi, and Kazunari Tanaka}



\acknowledgements{We acknowledge the support of the JSPS-CNRS
  bilateral program ``FoRmal tools for IoT sEcurity'' and the JSPS
  KAKENHI Grant Number 18H03204, thank the projects' participants, in
  particular Akira Tanaka for his comments on code extraction, and the
  anonymous reviewers for their comments that helped improve this paper.}

\supplement{The accompanying \coq{} development can be found on GitHub (see~\cite{compact}).}

\EventEditors{John Harrison, John O'Leary, and Andrew Tolmach}
\EventNoEds{3}
\EventLongTitle{10th International Conference on Interactive Theorem Proving (ITP 2019)}
\EventShortTitle{ITP 2019}
\EventAcronym{ITP}
\EventYear{2019}
\EventDate{September 9--12, 2019}
\EventLocation{Portland, OR, USA}
\EventLogo{}
\SeriesVolume{141}
\ArticleNo{28}

\begin{document}

\maketitle

\begin{abstract}
Succinct data structures give space-efficient representations of large
amounts of data without sacrificing performance. They rely
on cleverly designed data representations and algorithms.
We present here the formalization in Coq/SSReflect of two different
tree-based succinct representations and their accompanying algorithms. One is the Level-Order Unary Degree Sequence,
which encodes the structure of a tree in breadth-first
order as a sequence of bits, where access operations can be defined in
terms of Rank and Select, which work in constant time for static bit
sequences. The other represents dynamic bit sequences as binary
balanced trees, where Rank and Select present a low logarithmic
overhead compared to their static versions, and with efficient insertion
and deletion. The two can be stacked to provide a dynamic representation
of dictionaries for instance. While both representations are
well-known, we believe this to be their first formalization and
a needed step towards provably-safe implementations of big data.
\end{abstract}

\section{Introduction}

Succinct data structures~\cite{navarro2016} represent combinatorial objects (such as bit
vectors or trees) in a way that is space-efficient (using a number of
bits close to the information theoretic lower bound) and
time-efficient (i.e., not slower than classical algorithms). This
topic is attracting all the more attention as we are now collecting
and processing large amounts of data in various domains such as
genomes or text mining. As a matter of fact, succinct data
structures are now used in software products of data-centric companies
such as Google~\cite{kudo2011efficient}.

The more complicated a data structure is, the harder it is to process
it. A moment of thought is enough to understand that constant-time
access to bit representations of trees requires ingenuity. Succinct
data structures therefore make for intricate algorithms and their
importance in practice make them perfect targets for formal
verification~\cite{tanaka2016icfem}.

In this paper, we tackle the formal verification of tree algorithms
for succinct data structures. We first start by formalizing basic
operations such as counting (\rank{}) and searching (\select{}) bits
in arrays. This is an important step because
the theory of these basic operations sustains most succinct data
structures. Next, we formally define and verify a bit representation
of trees called Level-Order Unary Degree Sequence (hereafter LOUDS).  It is for example
used in the Mozc Japanese input method~\cite{kudo2011efficient}. The challenge there is that this
representation is based on a level-order (i.e., breadth-first) traversal
of the tree, which is difficult to describe in a structural way.
Nonetheless, like most succinct data
structures, this bit representation only deals with static data. Last,
we further explore the advanced topic of dynamic bit vectors. The implementation of the latter
requires to combine static bit vectors from succinct data structures
with classical balanced trees. We show in particular how this can be
formalized using a flavor of red-black trees where the data is in the
leaves (rather than in the internal nodes, as in most functional
implementations).

In both cases, our code can be seen as a verified functional
specification of the
algorithms involved. We were careful to use the right abstractions in
definitions so that this specification could be easily translated to
efficient code using arrays. For LOUDS we only rely on the \rank{} and
\select{} functions; we have already provided an efficient
implementation for \rank{}~\cite{tanaka2016icfem}.
For dynamic bit vectors, while the code we present here is functional,
it closely matches the algorithms given in~\cite{navarro2016}.
We did prove all the essential correctness properties, by showing the
equivalence of each operation with its functional counterpart
(functions on inductive trees for LOUDS, and on sequences of bits for
dynamic bit vectors).

Independently of this verified functional specification, we identify
two technical contributions, that arised while doing this
formalization.  One is the notion of level-order traversal up to a
path in a tree, which solves the challenge of performing
path-induction on a level-order traversal. Another is our experience
report with using small-scale reflection to prove algorithms on
inductive data, which we hope could provide insights to other
researchers.

The rest of this paper is organised as follows. The next section
introduces \rank{} and \select{}. Section~\ref{sec:louds} describes
our formalization of LOUDS, including the notion of level-order
traversal up to a path. Section~\ref{sec:dynamic_vectors} uses trees
to represent bit vectors, defining not only \rank{} and \select{}, but
also insertion and deletion. Section~\ref{sec:proof_techniques}
reports on our experience. Section~\ref{sec:related_work} compares
with the litterature, and Section~\ref{sec:conclusion} concludes.

\section{Two functions to build them all}
\label{sec:rank_and_select}

The \rank{} and \select{} functions are the most basic blocks to form
operations on succinct data structures: \rank{} counts bits while
\select{} searches for their position. The rest of this paper (in
particular Sect.~\ref{sec:louds_functions} and
Sect.~\ref{sec:dynamic_vectors}) explains how they are used in
practice to perform operations on trees.
In this section, we just briefly explain their formalization and
theory.

\subsection{Counting bits with \rank{}}
\label{sec:counting_rank}

\begin{figure}
\centering
\begin{tikzpicture}
\edef\mya{0}
\node (a0) at (0,0) {{\sl bitstring}};
\node (a1) at (40pt,0) {1001};
\node (a2)[right of=a1,xshift=-7pt] {0100};
\node (a3)[right of=a2,xshift=-7pt] {1110};
\node (a4)[right of=a3,xshift=-7pt] {0100};
\node (a5)[right of=a4] {1101};
\node (a6)[right of=a5,xshift=-7pt] {0000};
\node (a7)[right of=a6,xshift=-7pt] {1111};
\node (a8)[right of=a7,xshift=-7pt] {0100};
\node (a9)[right of=a8] {1001};
\node (a10)[right of=a9,xshift=-7pt] {1001};
\node (a11)[right of=a10,xshift=-7pt] {0100};
\node (a12)[right of=a11,xshift=-7pt] {0100};
\node (a13)[right of=a12] {0101};
\node (a14)[right of=a13,xshift=-7pt] {0101};
\node (a15)[right of=a14,xshift=-12pt] {10};
\foreach \from in {a1,a2,a3,a4,a5,a6,a7,a8,a9,a10,a11,a12,a13,a14,a15}
  {
   \draw[dotted] ($(\from.west)+(+0.1,0)$) -- +(0,-0.6);
   \node[anchor=west,xshift=-5pt] at ($(\from.west)+(+0.2,-0.5)$) {{\scriptsize \mya}};
   \pgfmathparse{int(\mya+4)}
   \xdef\mya{\pgfmathresult}
  }
\node[below of=a0,yshift=15pt] {{\scriptsize {\sl bit indices}}};
\draw[<->] ($(a1.north west)+(0,5pt)$) -- ($(a15.north east)+(0,5pt)$) ;
\node (lbl) at ($(a7.north)+(10pt,10pt)$) {length $n = 58$} ;

\draw[decoration={brace,mirror,aspect=0.50},decorate,thick,red] ($(a1.south west)+(0,-17pt)$)--($(a1.south east)+(0,-17pt)$) ;
\node at ($(a1.south)+(0,-25pt)$) {{\color{red}{$\rank(4) = 2$}}} ;
\node at ($(a1.south)+(0,-35pt)$) {{\color{red}{$\select(2) = 4$}}} ;

\draw[decoration={brace,mirror,aspect=0.55},decorate,thick,coqgreen] ($(a1.south west)+(0,-15pt)$)--($(a9.south east)+(0,-15pt)$) ;
\node at ($(a6.south)+(0,-24pt)$) {{\color{coqgreen}{$\rank(36) = 17$}}} ;
\node at ($(a6.south)+(0,-34pt)$) {{\color{coqgreen}{$\select(17) = 36$}}} ;

\draw[decoration={brace,mirror,aspect=0.75,raise=5pt},decorate,thick,blue] ($(a1.south west)+(0,-8pt)$)--($(a15.south east)+(0,-8pt)$) ;
\node at ($(a12.south)+(0,-23pt)$) {{\color{blue}{$\rank(58) = 26$}}} ;
\node at ($(a12.south)+(0,-33pt)$) {{\color{blue}{$\select(26) = 57$}}} ;
\end{tikzpicture}
\caption{Examples of \rank{} and \select{} queries on a sample bitstring (bit indexed from $0$ to $57$)}
\label{fig:rankselect}
\end{figure}
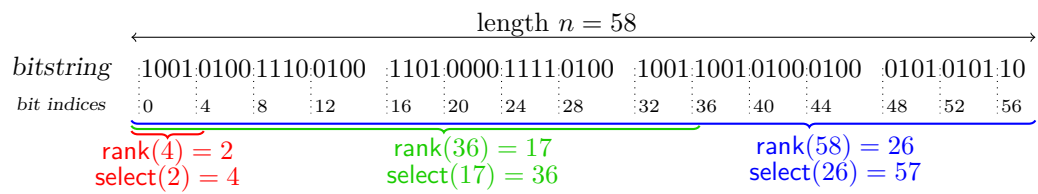

The \coqin{rank} function counts the number of elements~\coqin{b}
(most often bits) in the prefix (i.e., up to some index~\coqin{i}) of
an array~\coqin{s}. It can be conveniently formalized using
standard list functions:
\begin{minted}{ssr}
Definition rank b i s := count_mem b (take i s).
\end{minted}
Figure~\ref{fig:rankselect} provides several examples of \rank{}
queries.
The mathematically-inclined reader can alternatively\footnote{This is
  actually the definition that appears in Wikipedia at the time of
  this writing.} think of \rank{} as the cardinal of the number of indices
of \coqin{b}~bits in a tuple~\coqin{B}:
\begin{minted}{ssr}
Definition Rank (i : nat) (B : n.-tuple T) :=
  #|[set k : [1,n] | (k <= i) && (tacc B k == b)]|.
\end{minted}
In this definition,
\coqin{n.-tuple T} denotes%
\footnote{The notation \coqin{.-tuple} is a \ssreflect{} idiom for a
  suffix operator. Similarly we use \coqin{.+1} and \coqin{.-1} for
  successor and predecessor.}
sequences of \coqin{T} of length \coqin{n};
\coqin{[1,n]} is the type of integers between $1$ and $n$;
and \coqin{tacc} accesses the tuple counting the indices from~$1$.

\subsection{Finding bits with \select{}}
\label{sec:finding_select}

Intuitively, compared with \rank{}, \select{} performs the converse
operation: it returns the index of the \coqin{i}-th occurrence of
\coqin{b}, i.e., the {\em minimum\/} index whose rank is \coqin{i}.
It is conveniently specified using the \coqin{ex_minn} construct of the
\ssreflect{} library~\cite{ssrman}:
\begin{minted}{ssr}
Variables (T : eqType) (b : T) (n : nat).
Lemma select_spec (i : nat) (B : n.-tuple T) :
  exists k, ((k <= n) && (Rank b k B == i)) || (k == n.+1) && (count_mem b B < i).
Definition Select i (B : n.-tuple T) := ex_minn (select_spec i B).
\end{minted}
With this definition, \select{} returns the index of the sought bit
{\em plus one} (counting indices from~$0$); \select{}ing the
$0^{\rm th}$ bit always returns~$0$; when no adequate bit is found, \select{}
returns the size of the array plus one. The need for the $0$ case
explains why it makes sense to return indices starting from $1$.
Figure~\ref{fig:rankselect} provides several examples to illustrate
the \select{} function.

\subsection{The theory of \rank{} and \select{}}
\label{sec:theory_rank_select}

The \rank{} and \select{} functions are used in a variety of
applications whose formal verification naturally calls for a shared
library of lemmas.  Our first work is to identify and isolate this
theory. Its lemmas are not all difficult to prove. For instance, the
fact that \coqin{Rank} cancels \coqin{Select} directly follows from
the definitions:
\begin{minted}{ssr}
Lemma SelectK n (s : n.-tuple T) (j : nat) :
  j <= count_mem b s -> Rank b (Select b j s) s = j.
\end{minted}

However, as often with formalization, it requires a bit of work and
try-and-error to find out the right definitions and the right lemmas
to put in the theory of \rank{} and \select{}. For example, how
appealing the definition of \coqin{Select} above may be, proving its
equivalence with a functional version such as
\begin{minted}{ssr}
Fixpoint select i (s : seq T) : nat :=
  if i is i.+1 then
    if s is a :: s' then (if a == b then select i s' else select i.+1 s').+1
                    else 1
  else 0.
\end{minted}
turns out to add much comfort to the development of related lemmas.

As a consequence, the resulting theory of \rank{} and \select{}
sometimes looks technical and we therefore refer the reader to the
source code~\cite{compact} to better appreciate its current
status. Here, we just provide for the sake of completeness the
definition of two derived functions that are used later in this paper.

\subsubsection{The \succ{} and \pred{} functions}
\label{sec:succ}

In a bitstring, the \succ{} function computes the position of the next
$0$-bit or $1$-bit. It will find its use when dealing with LOUDS
operations in Sect.~\ref{sec:children}. More precisely, given a
bitstring \coqin{s}, \coqin{succ b s y} returns the index of the
next~\coqin{b} following index~\coqin{y}. This operation is
achieved by a combination of \rank{} and \select{}. First, a call to
\rank{} counts the number of \coqin{b}'s up to index~\coqin{y}; let
\coqin{N} be this number. Second, a call to \select{} searches for the
$\coqin{(N+1)}^{\rm th}$~\coqin{b}~\cite[p.~89]{navarro2016}:
\begin{minted}{ssr}
Definition succ (b : T) (s : seq T) y := select b (rank b y.-1 s).+1 s.
\end{minted}
In particular, there is no \coqin{b} in the set
$\{ \coqin{s}_i \,|\, \coqin{y} \leq i < \coqin{succ b s y} \}$:
\begin{minted}{ssr}
Lemma succP b n (s : n.-tuple T) (y : [1, n]) :
  b \notin \bigcup_(i : [1,n] | y <= i < succ b s y) [set tacc s i].
\end{minted}

Conversely, the \pred{} function computes the position of the previous bit and
will find its use in Sect.~\ref{sec:parent}. It is similar to
\succ{}, so that we only provide its definition for
reference:
\begin{minted}{ssr}
Definition pred (b : T) (s : seq T) y := select b (rank b y s) s.
\end{minted}

\section{LOUDS formalization}
\label{sec:louds}

\begin{figure}
\begin{subfigure}[t]{0.1\textwidth}
  \centering
  \begin{tikzpicture}[level distance=0.6cm,draw opacity=0,
    level 1/.style={sibling distance=1cm}]
    \node {level 0} 
      child { node {level 1} {
          child { node {level 2} {
            child { node {level 3} }
          } 
        }
      } } ;
  \end{tikzpicture}
\end{subfigure}
\begin{subfigure}[t]{0.44\textwidth}
  \centering
  \begin{tikzpicture}[level distance=0.6cm,
    level 1/.style={sibling distance=1cm}]
    \node {1}
    child { node[xshift=-2mm]{2}
      child { node[xshift=-2mm]{5} }
      child { node[xshift=2mm]{6} } }
    child { node {3} }
    child { node {4}
      child { node {7} }
      child { node {8}
        child {node {10} } } 
      child { node {9} } } ;
  \end{tikzpicture} \\
  {\footnotesize level-ordered enumeration of nodes: \vspace{1mm} \\
  $\begin{array}{|l|l|l|l|}
     \hline
     \mathrm{level}\;0 & \mathrm{level}\;1 & \mathrm{level}\;2 & \mathrm{level}\;3 \\
     \hline
     1 & 2,3,4 & 5,6,7,8,9 & 10 \\
     \hline
  \end{array}$}
  \caption{A sample tree}
  \label{fig:a_sample_tree}
\end{subfigure}
\begin{subfigure}[t]{0.44\textwidth}
  \centering
  \begin{tikzpicture}[level distance=0.6cm,
    level 1/.style={sibling distance=1cm}]
    \node {10}
    child { node {1110}
      child { node[xshift=-2mm]{110}
        child { node[xshift=-2mm]{0} }
        child { node[xshift=2mm]{0} } }
      child { node {0} }
      child { node {1110}
        child { node {0} }
        child { node {10}
          child {node {0} } } 
        child { node {0} } } } ;
  \end{tikzpicture} \\
  {\footnotesize Bitstring encoding: \vspace{0.2mm} \\
  $\begin{array}{|l|l|l|l|l|}
     \hline
     & \mathrm{level}\;0 & \mathrm{level}\;1 & \mathrm{level}\;2 & \mathrm{level}\;3 \\
     \hline
     10 & 1110 & 11001110 & 000100 & 0 \\
     \hline
  \end{array}$}
  \caption{Its LOUDS encoding}
\end{subfigure}
\caption{LOUDS encoding of a sample unlabeled tree}
\label{fig:louds}
\end{figure}
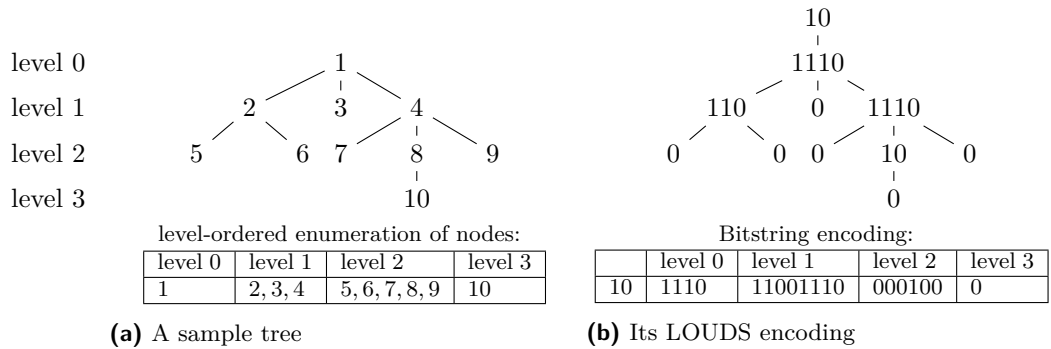

Operationally, a LOUDS encoding
consists in turning a tree into an array of bits via a level-order
traversal.
Figure~\ref{fig:louds} provides a concrete example. The
resulting array is the ordered concatenation of the bit representation
of each node. Each node is represented by a list of bits that contains
as many $1$-bits as there are children and that is terminated by a
$0$-bit.

The significance of the LOUDS encoding is that it preserves the
branching structure of the tree without pointers, making
for a compact representation in memory. Moreover, read-only
operations can be implemented using \rank{} and \select{},
which can be implemented in constant-time.

We explain how we formalize the LOUDS encoding in
Sect.~\ref{sec:louds_formalization} and how we formally verify the
correctness of operations on trees built out of \rank{} and \select{}
in Sect.~\ref{sec:louds_functions}.

\subsection{LOUDS encoding formalized in \coq{}}
\label{sec:louds_formalization}

We define arbitrarily-branching trees by an inductive type:
\begin{minted}{ssr}
Variable A : Type.
Inductive tree := Node : A -> seq tree -> tree.
Definition forest := seq tree.
Definition children_of_node : tree -> forest := ...
Definition children_of_forest : forest -> forest := flatten \o map children_of_node.
\end{minted}
where \coqin{\o} is the function composition operator (i.e., $\circ$),
and where the type \coqin{A} is the type of labels.
We also introduce the abbreviation \coqin{forest} for a list of trees,
and functions to obtain children.  With this definition of trees, a
leaf is a node with an empty list of children.
For example, the tree of Fig.~\ref{fig:louds} becomes in \coq{}:
\begin{minted}{ssr}
Definition t : tree nat := Node 1
  [:: Node 2 [:: Node 5 [::]; Node 6 [::]];
      Node 3 [::];
      Node 4 [:: Node 7 [::];
                 Node 8 [:: Node 10 [::]];
                 Node 9 [::]]].
\end{minted}

\subsubsection{Height-recursive level-order traversal}
\label{sec:lo_traversal}

The intuitive definition of level-order traversal
iterates on a forest, returning first the toplevel nodes of the
forest, then their children (applying \coqin{children_of_forest}), etc.
We parameterize the definition with an arbitrary function \coqin{f}
for generality.
\begin{minted}{ssr}
Variables (A B : Type) (f : tree A -> B).
Fixpoint lo_traversal' n (s : forest A) :=
  if n is n'.+1 then map f s ++ lo_traversal' n' (children_of_forest s) else [::].
Definition lo_traversal t := lo_traversal' (height t) [:: t].
\end{minted}
The parameter \coqin{n} is filled here with the maximum
height of the forest, meaning that we iterate just the right
number of times for the forest to become empty.

Yet, this definition is not fully satisfactory.
One reason is that it is not structural: we are not recursing on a
tree, but iterating on a forest, using its height as recursion index.
Another one is that, as we will see in
Sect.~\ref{sec:louds_functions}, the name {\em level-order} is
misleading. For many proofs, we are not interested in complete
traversal of the tree, level by level, but rather by partial
traversal along a path in the tree, where the forest we consider
actually overlaps levels.

\subsubsection{A structural level-order traversal}
\label{sec:structural_lo_traversal}

At first it may seem that the non-structurality is inherent to
level-order traversal. There is no clear way to build the sequence
corresponding to the traversal of a tree from those of its children.
However, Gibbons and Jones~\cite{Gibbons91,Jones93} showed that
this can be achieved by splitting the output into a list of levels.
One can combine two such structured traversals
by {\em zipping} them, i.e., concatenating corresponding levels,
and recover the usual traversal by flattening the list.
Since concatenation of lists forms a monoid, zipping of traversals
also forms a monoid.
\begin{minted}{ssr}
Variable (A : Type) (e : A) (M : Monoid.law e).
Fixpoint mzip (l r : seq A) : seq A := match l, r with
  | (l1::ls), (r1::rs) => (M l1 r1) :: mzip ls rs
  | nil, s | s, nil    => s
  end.

Lemma mzipA : associative mzip.
Lemma mzip1s s : mzip [::] s = s.
Lemma mzips1 s : mzip s [::] = s.
Canonical mzip_monoid := Monoid.Law mzipA mzip1s mzips1.
\end{minted}
Here \coqin{Monoid.Law}, from the \coqin{bigop} module of \ssreflect,
denotes an operator together with its neutral element (here
\coqin{[::]}) and the required monoidal equations, which are also
satisfied by~\coqin{mzip}.

We now define our traversal by instantiating \coqin{mzip} to the
concatenation monoid. The resulting \coqin{mzip_cat} is a structure of
type \coqin{Monoid.law [::]} that can be used as an operator of type
\coqin{seq (seq B) -> seq (seq B)} \coqin{-> seq (seq B)} enjoying the
properties of a monoid.
\begin{minted}{ssr}
Variables (A : eqType) (B : Type) (f : tree A -> B).
Definition mzip_cat := mzip_monoid (cat_monoid B).

Fixpoint level_traversal t :=
  [:: f t] :: foldr (mzip_cat \o level_traversal) nil (children_of_node t).

Lemma level_traversalE t :
  level_traversal t =
  [:: f t] :: \big[mzip_cat/nil]_(i <- children_of_node t) level_traversal i.

Definition lo_traversal_st t := flatten (level_traversal t).
Theorem lo_traversal_stE t : lo_traversal_st t = lo_traversal f t.
\end{minted}
To let \coq{} recognize the structural recursion, we have to use the
recursor \coqin{foldr} in the definition of \coqin{level_traversal}.
Yet, the intended equation is the one expressed by
\coqin{level_traversalE}, i.e., first output the image of the
node, and then combine the traversals of the children.
Then \coqin{lo_traversal_st}
can be proved equal to the previously defined
\coqin{lo_traversal}.
Deforestation can furthermore improve the efficiency of
\coqin{level_traversal}%
\footnote{See \coqin{level_traversal_cat} in \cite[\coqin{tree_traversal.v}]{compact}.}.

\subsubsection{LOUDS encoding}
\label{sec:louds_encoding}

Finally, the LOUDS encoding is obtained by instantiating
\coqin{lo_traversal_st} with an appropriate function (called the
\mydef{node description} of a node), and flattening once more:
\begin{minted}{ssr}
Definition node_description s := rcons (nseq (size s) true) false.
Definition children_description t := node_description (children_of_node t).
Definition LOUDS t := flatten (lo_traversal_st children_description t).
\end{minted}
Here, \coqin{rcons s x} adds \coqin{x} to the end of the sequence \coqin{s}, while \coqin{nseq n x}
creates a sequence consisting of \coqin{n} copies of \coqin{x}.
Note that we chose here not to add the usual ``{\tt 10}''
prefix~\cite[p.~212]{navarro2016} shown in Fig.~\ref{fig:louds}, as it
appeared to just complicate definitions. It can be easily recovered by
adding an extra root node, as ``{\tt 10}'' is the representation of a
node with 1 child.

For example, we can recover the encoding displayed in
Fig.~\ref{fig:louds} with this definition of \coqin{LOUDS}:\hspace{-2ex}
\begin{minted}{ssr}
Lemma LOUDS_t : LOUDS (Node 0 [:: t]) =
  [:: true; false; true; true; true; false;
      true; true; false; false; true; true; true; false;
      false; false; false; true; false; false; false].
\end{minted}

We can also prove some properties of this representation, such
as its size:
\begin{minted}{ssr}
Lemma size_LOUDS t : size (LOUDS t) = 2 * number_of_nodes t - 1.
\end{minted}
This is an easy induction, remarking that \coqin{size}\,\coqin{\o}\,\coqin{flatten}\,\coqin{\o}\,\coqin{flatten} is a morphism between \coqin{mzip_cat} and \coqin{+}.

\subsection{LOUDS functions using \rank{} and \select{}}
\label{sec:louds_functions}

In this section, we formalize LOUDS functions and prove their
correctness. These functions are essentially built out of \rank{} and
\select{}. Their correctness statements establish a correspondence
between operations on trees defined inductively and operations on
their LOUDS encoding.
We start by explaining how we represent positions in trees and then
comment on the formal verification of LOUDS operations using
representative examples.

\subsubsection{Positions in trees}
\label{sec:positioning}

For a tree defined inductively, we represent the position of a node as
usual: using a \mydef{path}, i.e., a list that records the branches
taken from the root to reach the node. For example, the position of
the node~$8$ in Fig.~\ref{fig:a_sample_tree} is \coqin{[:: 2; 1]}. Not
all positions are valid; we sort out the valid ones by means of the
predicate \coqin{valid_position} (definition omitted for brevity).

\begin{figure}
\centering
\hspace{8ex}
\begin{tikzpicture}
\fill[lightgray] (0,0) -- (1.8,-1.8) -- (0.6,-1.8) -- (0.6,-2.2) -- (-2.2,-2.2) -- cycle ;
\draw[black,fill=coqgreen] (0.6,-1.8) -- (1.8,-1.8) -- (2.2,-2.2) -- (0.6,-2.2) -- cycle;
\draw[black,fill=coqgreen] (0.6,-2.2) -- (-2.2,-2.2) -- (-2.6,-2.6) -- (0.6,-2.6) -- cycle;
\draw (0,0) -- (-3,-3) -- (3,-3) -- cycle ;
\draw[red,thick] (0,0) -- (0,-0.6) -- (0.6,-1.2) -- (0.6,-1.8);

\draw[fill=lightgray] (2.5,0) rectangle (3,-0.2) node[anchor=west,yshift=4pt] {~traversed tree} ;
\draw[fill=coqgreen] (2.5,-0.5) rectangle (3,-0.7) node[anchor=west,yshift=3pt]{~fringe} ;
\end{tikzpicture}
\caption{Level-order traversal of a tree up to a path}
\label{fig:generating}
\end{figure}
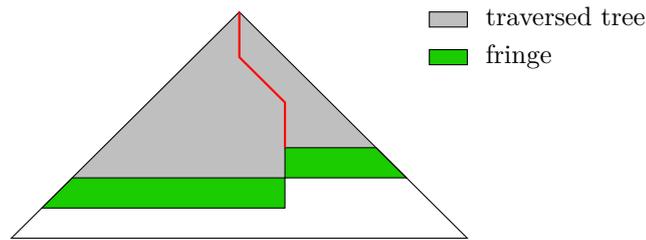

In contrast, the position of nodes in the LOUDS encoding is not
immediate.
We define it as the length of the generated LOUDS up to the
corresponding path. To do that, we first need to define a notion of
level-order traversal up to a path, which collects all the nodes
preceding the one referred by that path (which need not be valid):
\begin{minted}{ssr}
Definition split {T} n (s : seq T) := (take n s, drop n s).
Variables (A : eqType) (B : Type) (f : tree A -> B).
Fixpoint lo_traversal_lt (s : forest A) (p : seq nat) : seq B := match p, s with
  | nil, _ | _, nil => nil
  | n :: p', t :: s' =>
    let (fs, ls) := split n (children_of_node t) in
    map f (s ++ fs) ++ lo_traversal_lt (ls ++ children_of_forest (s' ++ fs)) p'
  end.
\end{minted}
This new traversal appears to be the key to clean proofs of LOUDS
properties.  In a previous
attempt using the height-recursive level-order traversal of
Sect.~\ref{sec:lo_traversal}, proofs were unwieldy (one needed to
manually set up inductions) and lemmas did not arise naturally.  We
expect this new traversal to have applications to other uses of
level-order traversal.

This definition may seem scary, but it closely corresponds to
the imperative version of level-order traversal, which relies on a
queue: to get the next node, take it from the front
of the queue, and add its children to the back of the queue. We
define our traversal so that the node we have reached is the one at
the front of the queue~\coqin{s}. To move to its $n^{\rm th}$ child
(indices starting from 0), we first output all the nodes in the queue,
and its children up to the previous one, and proceed with a new queue
containing the remaining children (starting from the $n^{\rm th}$) and the
children of the other nodes we have just output.
Figure~\ref{fig:generating} shows how the traversal progresses. The
point is that as soon as the queue spans
all the fringe of the traversed tree, it is able to generate the
remainder of the traversal.
We can verify that \coqin{lo_traversal_lt} indeed qualifies as a level-order
traversal by proving that its
output converges to the full level-order traversal
when the length of \coqin{p} reaches the height of the tree:
\begin{minted}{ssr}
Theorem lo_traversal_ltE (t : tree A) (p : seq nat) :
  size p >= height t -> lo_traversal_lt [:: t] p = lo_traversal_st f t.
\end{minted}
We also introduce a function that computes the fringe of the
traversal up to \coqin{p}, i.e., the forest generating the
remainder of the traversal.
\begin{minted}{ssr}
Fixpoint lo_fringe (s : forest A) (p : seq nat) : forest A := ...
Lemma lo_traversal_lt_cat s p1 p2 :
  lo_traversal_lt s (p1 ++ p2) =
  lo_traversal_lt s p1 ++ lo_traversal_lt (lo_fringe s p1) p2.
\end{minted}
We omit the definition but the lemma states exactly
this property. It decomposes the traversal generated by a path,
allowing induction from either end of the list representing the position.

Using the path-indexed traversal function, we can directly obtain the
index of a node in the level-order traversal of a tree:
\begin{minted}{ssr}
Definition lo_index (s : forest A) (p : seq nat) := size (lo_traversal_lt id s p).
\end{minted}
The expression \coqin{lo_index [:: t] p} counts the number of nodes in the
traversal of \coqin{t} before the position~\coqin{p}.
Similarly, we give an alternative definition of the LOUDS encoding, and use
it to map a position in the tree to a position in its encoding
(i.e., the index of the first bit of the representation of a node):
\begin{minted}{ssr}
Definition LOUDS_lt s p := flatten (lo_traversal_lt children_description s p).
Definition LOUDS_position s p := size (LOUDS_lt s p).
\end{minted}
Here the position in the whole tree is obtained
as \coqin{LOUDS_position [:: t] p}, but we can also compute relative
positions by using \coqin{LOUDS_position s p} where \coqin{s} is a
generating forest whose front node is the one we start from.
Note that both \coqin{lo_index} and \coqin{LOUDS_position} return
indices starting from 0.

For example, in Fig.~\ref{fig:louds}, the position of the node~$8$ is
\coqin{[:: 2; 1]} in the inductively defined tree and \coqin{17} in
the LOUDS encoding:
\begin{minted}{ssr}
Definition p8 := [:: 2; 1].
Eval compute in LOUDS_position [:: Node 0 [:: t]] (0 :: p8). (* 17 *)
\end{minted}

Finally, here is one of the essential lemmas for proofs on LOUDS,
which relates \coqin{lo_index} and \coqin{LOUDS_position} using
\coqin{select}:
\begin{minted}{ssr}
Lemma LOUDS_position_select s p p' : valid_position (head dummy s) p ->
  LOUDS_position s p = select false (lo_index s p) (LOUDS_lt s (p ++ p')).
\end{minted}
Namely if the index of \coqin{p} is $n$, then its position in the
LOUDS encoding is the index of its $n^{\rm th}$ 0-bit (recall that
\coqin{select} counts indices starting from 1).
Here \coqin{p'} allows us to complete \coqin{p} to a path of
sufficient length, so that \coqin{LOUDS_lt} converges to \coqin{LOUDS}.

\subsubsection{Number of children using \succ{}}
\label{sec:children}

As a first example, let use formalize the LOUDS function that
counts the number of children of a node. For a tree defined
inductively, this operation can be achieved by first walking down the
path to the node and then looking at the list of its children.
\begin{minted}{ssr}
Fixpoint subtree (t : tree) (p : seq nat) :=
  if p is n :: p' then subtree (nth t (children_of_node t) n) p' else t.
Definition children t p := size (children_of_node (subtree t p)).
\end{minted}

To count the number of children of a node using a LOUDS encoding, one
first has to notice that each node is
terminated by a $0$-bit. Given such a $0$-bit (or equivalently the
corresponding node), one can find the number of children by
computing the distance with the next $0$-bit
\cite[p.~214]{navarro2016}. Finding this bit is the purpose
of the \succ{} function of Sect.~\ref{sec:succ}:
\begin{minted}{ssr}
Definition LOUDS_children (B : bitseq) (v : nat) : nat := succ false B v.+1 - v.+1.
\end{minted}
The \coqin{.+1} offset comes from the fact \coqin{succ} computes on
indices starting from 1.

\coqin{LOUDS_children} is correct because, when applied to the
\coqin{LOUDS_position} of a position~\coqin{p}, it produces the same
result as the function~\coqin{children}:
\begin{minted}{ssr}
Theorem LOUDS_childrenE (t : tree A) (p p' : seq nat) :
  let B := LOUDS_lt [:: t] (p ++ 0 :: p') in
  valid_position t p -> LOUDS_children B (LOUDS_position [:: t] p) = children t p.
\end{minted}

\subsubsection{Parent and child node using \rank{} and \select{}}
\label{sec:parent}

A path in a tree defined inductively gives direct ancestry
information. In particular, removing the last index denotes the
parent, and adding an extra index denotes the corresponding
child. It takes more ingenuity to find parent and child using a LOUDS
representation and functions from Sect.~\ref{sec:rank_and_select}
alone. The idea is to count the number of nodes and branches up to the
position in question~\cite[p.~215]{navarro2016}. More precisely, given a LOUDS
position~\coqin{v}, let \coqin{Nv} be the number of nodes up
to~\coqin{v} (\coqin{rank false v B} computes this number).
Then, \coqin{select true Nv B} looks for the \coqin{Nv}-th down-branch,
which is the branch leading to the node of position~\coqin{v}.
Last, this branch belongs to a node whose position can be recovered
using the \coqin{pred} function (from Sect.~\ref{sec:succ}).
Reciprocally, one computes the $i^{\rm th}$ child by using \coqin{rank
  true} and \coqin{select false}.
This leads to the following definitions:
\begin{minted}{ssr}
Definition LOUDS_parent (B : bitseq) (v : nat) : nat :=
  let j := select true (rank false v B) B in pred false B j.
Definition LOUDS_child (B : bitseq) (v i : nat) : nat :=
  select false (rank true (v + i) B).+1 B.
\end{minted}

One can check the correctness of \coqin{LOUDS_parent} and
\coqin{LOUDS_child} as follows.
Consider a node reached by the path~\coqin{rcons p i}. Its parent is
the node reached by the path~\coqin{p}, and conversely it is the
$i^{\rm th}$ child of this node. We can formally prove that the
LOUDS position of~\coqin{p} (respectively \coqin{rcons p i}) and
the position computed by \coqin{LOUDS_parent} (respectively
\coqin{LOUDS_child}) coincide:
\begin{minted}{ssr}
Variables (t : tree A) (p p' : seq nat) (i : nat).
Hypothesis HV : valid_position t (rcons p i).
Let B := LOUDS_lt [:: t] (rcons p i ++ p').
Theorem LOUDS_parentE :
  LOUDS_parent B (LOUDS_position [:: t] (rcons p i)) = LOUDS_position [:: t] p.
Theorem LOUDS_childE :
  LOUDS_child B (LOUDS_position [:: t] p) i = LOUDS_position [:: t] (rcons p i).
\end{minted}

\smallskip

The approach that we explained so far shows how to carry out the
formal verification of the LOUDS operations that are listed
in~\cite[Table~8.1]{navarro2016}. However, how useful they may be
for many big-data applications, these operations assume static compact
data structures. The next section
explains how to extend our approach to deal with dynamic structures.

\section{Dynamic bit vectors}
\label{sec:dynamic_vectors}

In some applications bit vectors
need to support dynamic operations---not just static queries.
We formalize such \mydef{dynamic bit vectors}, and implement and
verify
``dynamic operations'' on them: inserting a bit into a
bit vector, and deleting a bit from one.

In Sect.~\ref{sec:representing_dynamic_vectors}, we explain the data
structure that allows for an efficient implementation of dynamic
operations.
In Sect.~\ref{sec:basic_queries}, we formalize the \rank{} and
\select{} queries.
Sections~\ref{sec:insert} and~\ref{sec:delete} are dedicated to the
formalization of the more difficult insertion and deletion.

\subsection{Representing dynamic bit vectors}
\label{sec:representing_dynamic_vectors}

The choice of representation for dynamic bit vectors is motivated by
complexity considerations.
Insertion into a linear array has time complexity $O(n)$, but we can
improve this by using a balanced binary search tree to represent the
bit array, which enables us to handle insertions in at most
$O(w)$ time, with a trade-off of $O(n/w)$ bits of extra space,
where $w$ is a parameter controlling the width of
each tree node and should no more than the size of a
native machine word in bits\footnote{The complexity bounds referred to in this
  section are dependent on the model of computation used. Here, we
  assume that we are working with a sequential RAM machine, where we have
  $O(w) = O(\log n)$ as we can only address at most $2^w$ bits of memory.}~\cite{navarro2016}:
i.e., for a typical 64-bit machine, we would set $w$ to 32 or 64.

On a side note, balanced binary trees are certainly not the most compact data structure that could be used here.
In fact, various data structures with better complexity have been designed~\cite{navsad14, ramanetal01},
however those structures are complicated and are unlikely to offer practical improvements over the structure presented
here~\cite{navarro2016}. As a result, we choose to work only with balanced binary trees, which are much easier to reason about.

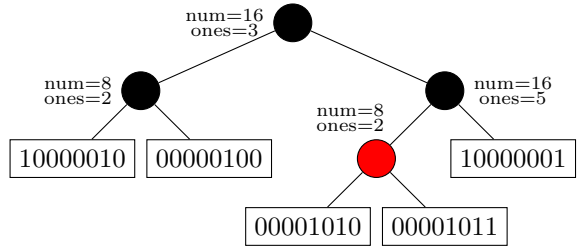
\begin{figure}
\centering
\begin{tikzpicture}
[every node/.style={draw,level distance=8mm},level 1/.style={sibling distance=40mm,level distance=9mm},level 2/.style={sibling distance=18mm}]
\newcommand\metadata[2]{$\substack{\text{num} = #1 \\ \text{ones} = #2}$}
\begin{scope}[minimum size=5mm]
\node[fill=black, circle, label=180:\metadata{16}{3}] {}
child { node[fill=black, circle, label=180:\metadata{8}{2}] {} child { node{10000010} } child { node{00000100} } }
child { node[fill=black, circle, label=0:\metadata{16}{5}] {} 
 child { node[fill=red, circle, label={[xshift=-1cm]70:\metadata{8}{2}}] {} child { node{00001010} } child { node{00001011} } }
 child { node{10000001} } };
\end{scope}
\end{tikzpicture}
\caption{Example of tree representation of a dynamic bit vector}
\label{fig:dynamicbitvector}
\end{figure}

In our formalization of the dynamic bit vector's %insertion and deletion
algorithms, we use a red-black tree as our balanced tree structure.
Each node holds a color and meta-data about the bit vector, and each
leaf holds a \mydef{flat\/} (i.e., list-based) bit array. Following Navarro~\cite{navarro2016}, we store two
natural numbers in each node: the size and the rank of the left
subtree (recorded as ``num'' and ``ones'' in
Fig.~\ref{fig:dynamicbitvector}).
\begin{minted}{ssr}
Inductive color := Red | Black.
Inductive btree (D A : Type) : Type :=
 | Bnode of color & btree D A & D & btree D A
 | Bleaf of A.

Definition dtree := btree (nat * nat) (seq bool).
\end{minted}

Our first step is to formalize the structural invariant of our tree
representation of bit vectors, which is required to prove the
correctness of queries and updates on it.
It states that the numbers encoded in each node are
the left child's size and rank, and that leaves contain a number of
bits between \coqin{low} and \coqin{high}.
\begin{minted}{ssr}
Variables low high : nat.  (* instantiated as w^2 / 2 and w^2 * 2 *)
Fixpoint wf_dtree (B : dtree) := match B with
  | Bnode _ l (num, ones) r => [&& num == size (dflatten l),
                                   ones == count_mem true (dflatten l),
                                   wf_dtree l & wf_dtree r]
  | Bleaf arr               => low <= size arr < high
  end.
\end{minted}
Here, the function \coqin{dflatten} defines the semantics of our
tree representation of a bit vector (\coqin{dtree}) by converting it
to a flat representation of that vector:
\begin{minted}{ssr}
Fixpoint dflatten (B : dtree) := match B with
  | Bnode _ l _ r => dflatten l ++ dflatten r
  | Bleaf s => s
  end.
\end{minted}

\subsection{Verifying basic queries}
\label{sec:basic_queries}

The basic query operations can be easily defined via traversal of the
tree. We implement the queries \rank{}, $\select_{1}$, and
$\select_{0}$ as the \coq{} functions \coqin{drank},
\coqin{dselect_1}, and \coqin{dselect_0}.  For example, \coqin{drank}
is implemented as follows, using the (static) \coqin{rank} function
from Sect.~\ref{sec:counting_rank}:
\begin{minted}{ssr}
Fixpoint drank (B : dtree) (i : nat) := match B with
  | Bnode _ l (num, ones) r =>
    if i < num then drank l i else ones + drank r (i - num)
  | Bleaf s => rank true i s
  end.
\end{minted}

We prove that our function \coqin{drank} indeed computes the query
\rank{} using a custom induction principle \coqin{dtree_ind},
corresponding to the predicate \coqin{wf_dtree}:
\begin{minted}{ssr}
Lemma drankE (B : dtree) i : wf_dtree B -> drank B i = rank true i (dflatten B).
Proof. move=> wf; move: B wf i. apply: dtree_ind. (* ... *) Qed.
\end{minted}
Note that our implementation is only correct on well-formed trees.

The formalization and verification of the \select{} queries proceed
along the same lines.

\subsection{Implementing and verifying insertion}
\label{sec:insert}

Insertion is significantly harder to implement than static
queries. We need to maintain the invariant on the size of the
leaves, which means that we have to split a leaf if it becomes too big,
and in that case we may need to rebalance the tree, to maintain the
red-black invariant, updating the meta-data on the way.

We translate the algorithm given by Navarro~\cite{navarro2016}
directly into \coq{}. Here, \coqin{high} is the maximum number of bits a leaf
can contain before it needs to be split up:
\begin{minted}{ssr}
Definition dins_leaf s b i :=
  let s' := insert1 s b i in (* insert element b in sequence s at position i *)
  if size s + 1 == high then
    let n := size s' %/ 2 in let sl := take n s' in let sr := drop n s' in
    Bnode Red (Bleaf _ sl) (n, count_mem true sl) (Bleaf _ sr)
  else Bleaf _ s'.

Fixpoint dins (B : dtree) b i : dtree := match B with
  | Bleaf s => dins_leaf s b i
  | Bnode c l d r =>
      if i < d.1 then balanceL c (dins l b i) r (d.1.+1, d.2 + b)
                 else balanceR c l (dins r b (i - d.1)) d
  end.

Definition dinsert (B : btree D A) b i : btree D A :=
  match dins B b i with
  | Bleaf s => Bleaf _ s
  | Bnode _ l d r => Bnode Black l d r
  end.
\end{minted}
\coqin{dins} recurses on the tree, searching for the leaf where the
insertion must be done, calling then \coqin{dins_leaf}, which inserts
a bit in the leaf, eventually splitting it if required. On its way back,
\coqin{dins} calls balancing functions \coqin{balanceL} and
\coqin{balanceR} to maintain the red-black invariant.
We omit the code of the balancing functions (see~\cite{compact}).
Like the standard version, they fix imbalances possibly occurring on
the left and on the right, respectively, but they must also adjust the
meta-data in the nodes.
\coqin{dinsert} is a simple wrapper over \coqin{dins} that completes
the insertion by painting the root black.
The real definitions are more abstract~\cite{compact};
we chose to instantiate them in this paper for readability.
%For readability, we have inlined some instantiations in this paper; the
%real definitions~\cite{compact} are more abstract.
% In general, the implementation is rather simple at less than 40 lines.

\def\preservedata{(a)}
\def\maintainproperties{(b)}
\def\returnbalanced{(c)}

Verifying \coqin{dinsert} requires verifying three different
properties: \coqin{dinsert} must \preservedata{}~preserve the data,
\maintainproperties{}~maintain the structural invariants of the
tree, and
\returnbalanced{}~return a balanced red-black tree.
Properties~\preservedata{} and~\maintainproperties{} are related, in
that the latter is required by the former.
\begin{minted}{ssr}
Notation wf_dtree_l := (wf_dtree low high).
Definition wf_dtree' t := if t is Bleaf s then size s < high else wf_dtree_l t.
Lemma wf_dtree_dtree' t : wf_dtree_l t -> wf_dtree' t.
Lemma wf_dtree'_dtree t : wf_dtree' t -> wf_dtree 0 high t.

Lemma dinsertE (B : dtree) b i :
  wf_dtree' B -> dflatten (dinsert B b i) = insert1 (dflatten B) b i.
Lemma dinsert_wf (B : dtree) b i : wf_dtree' B -> wf_dtree' (dinsert B b i).
\end{minted}
% Lemma dinsE (B : dtree) b i :
%   wf_dtree_l B -> dflatten (dins B b i) = insert1 (dflatten B) b i.
% Lemma dins_wf (B : dtree) b i : wf_dtree_l B -> wf_dtree_l (dins B b i).
A subtle point here is that we may start from a tree formed of a
single small leaf, i.e., a leaf smaller than \coqin{low}.
To handle this situation we introduce \coqin{wf_dtree'}, which does
not enforce the lower bound on this single leaf. This new predicate is
entailed by the original invariant (it removes one check), but
interestingly it also entails it if we set the lower bound to 0. Since
the queries of Sect.~\ref{sec:basic_queries} were proved with abstract
lower and upper bounds,
their proofs are readily usable through this weakening. However, we need to
use \coqin{wf_dtree'} when we prove properties of \coqin{dinsert}, as
it modifies the tree.

Proving~\preservedata{} and~\maintainproperties{} involves no theoretical difficulty. We explain
in Sect.~\ref{sec:proof_techniques} some techniques to
write short proofs: about 100 lines in total for both properties,
including lemmas for \coqin{balanceL} and \coqin{balanceR}, which
involve large case analyses.

Property~\returnbalanced{} about \coqin{dinsert} never breaking the
red-black tree invariant is notoriously more challenging.  More
importantly, we want to eliminate cases where the ``height balance''
at a node is broken. It is easy to model the property that no red node
has a red child; the ``height balance'' property is modeled using the
black-depth. We can thus model the red-black tree invariant with a
recursive function that takes as arguments the ``color context''
\coqin{ctxt} (the color of the parent's node) and the black-depth of
the node \coqin{bh}:
\begin{minted}{ssr}
Fixpoint is_redblack (B : dtree) (ctxt : color) (bh : nat) := match B with
  | Bleaf _ => bh == 0
  | Bnode c l _ r => match c, ctxt with
    | Red, Red   => false
    | Red, Black => is_redblack l Red bh && is_redblack r Red bh
    | Black, _   => (bh > 0) && is_redblack l Black bh.-1
                             && is_redblack r Black bh.-1
  end end.
\end{minted}

To show that \coqin{dinsert} preserves the red-black tree property, we
define and prove a number of weaker structural lemmas that are
basically equivalent to stating that a tree returned by \coqin{dins}
is structurally valid if the root is painted black.  We do not
describe the proof in detail because the technique is
well-known~\cite{okasaki98} and has been formalized in multiple
sources (see Sect.~\ref{sec:related_work}). Using these weaker lemmas,
we can prove the following structural validity lemma:
\begin{minted}{ssr}
Lemma dinsert_is_redblack (B : dtree) b i n :
  is_redblack B Red n -> exists n', is_redblack (dinsert B b i) Red n'.
\end{minted}

\subsection{Deletion: searching for invariants}
\label{sec:delete}

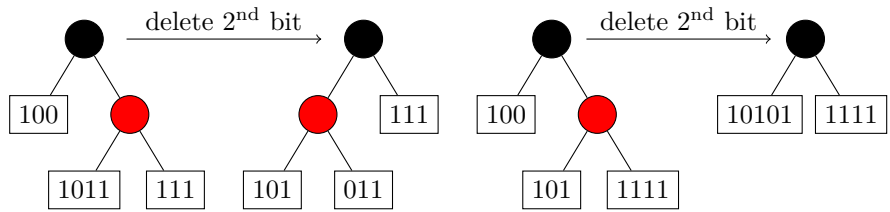
\begin{figure}
\centering
\begin{tikzpicture}
[every node/.style={draw,sibling distance=12mm,level distance=10mm}]
\begin{scope}[minimum size=5mm]
\matrix[draw=none, column sep=5mm]
{
\node[fill=black, circle] (LHS5) {} child { node{100} } child { node[fill=red, circle] {} child { node{1011} } child { node{111} } };
& \node[fill=black, circle] (RHS5) {} child { node[fill=red, circle]
  {} child { node{101} } child { node{011} } }  child { node {111} };
& \node[fill=black, circle] (LHS4) {} child { node{100} } child { node[fill=red, circle] {} child { node{101} } child { node{1111} } };
& \node[fill=black, circle] (RHS4) {} child {
  node{10101} } child { node{1111} }; \\
% & \node[fill=black, circle, label=\coqin{Stay}] (RHS5) {} child { node{10001} } child { node{1111} }; \\
};
\end{scope}
\draw[->, shorten <= 2mm, shorten >= 2mm] (LHS4) -- (RHS4) node[midway, above, draw=none] {delete 2${}^\mathrm{nd}$ bit};
\draw[->, shorten <= 3mm, shorten >= 3mm] (LHS5) -- (RHS5) node[midway, above, draw=none] {delete 2${}^\mathrm{nd}$ bit};
\end{tikzpicture}
\caption{Base cases of \delete{} ($\text{lower bound} = 3$) }
\label{fig:deletion}
\end{figure}

Deletion in dynamic bit vectors is difficult for two reasons.
One is that, in order to maintain the upper and lower
bounds on the size of leaves, which is required to attain
simultaneously space and time efficiency, deleting a bit in a leaf may
require some rearrangement of the surrounding nodes.
Figure~\ref{fig:deletion} shows the result of deleting a bit in a leaf
of the tree, when this leaf has already the smallest allowed size.
This can be resolved by borrowing a bit from a sibling (left case), or
merging two siblings (right case), but depending on the configurations
of nodes, this may require to first rotate the tree.

The other is that deletion in a functional red-black tree is
a complex operation~\cite{kahrs01}, and that finding how to adapt
the invariants of the litterature to our specific case proved to be
non-trivial.
Therefore, we took a twofold approach. First, we searched for invariants
in a concrete tree structure with invariants encoded using
dependent types.
Then, we removed dependent types and implemented \delete{}
and proved its correctness (more details in Sect.~\ref{sec:proof_techniques}).

Contrary to insertion, knowing the color of the modified child is not
sufficient to rebalance its parent correctly after deletion, and
recompute its meta-data. We need to propagate two more pieces of
information: whether the black-height decreased (\coqin{d_down}
below), and the meta-data corresponding to the deleted bit
(\coqin{d_del}). We encapsulate these in a ``tree state'':
\begin{minted}{ssr}
Record deleted_dtree: Type := MkD { d_tree :> dtree; d_down: bool; d_del: nat*nat }.
\end{minted}
Note that \coqin{deleted_dtree} is automatically coerced to \coqin{dtree}.

Now, we can define \delete{} in the natural way, but we need to take
care about balance operations and invariants on the size of leaves.
Specifically, the balance operations must be reimplemented as \coqin{balanceL'} and \coqin{balanceR'}, 
which need to satisfy the following invariants, i.e., the resulting ``balanced'' tree is \emph{deleted-red-black}
(i.e., a red-black tree, either with the same black height, or with a black
root and decreased black height), given that the unproblematic subtree
is red-black, while the unbalanced one is deleted-red-black.

\begin{minted}{ssr}
Definition balanceL' (c:color)(l:deleted_dtree)(d:nat*nat)(r:dtree):deleted_dtree :=
Definition balanceR' (c:color)(l:dtree)(d:nat*nat)(r:deleted_dtree):deleted_dtree :=

Definition is_deleted_redblack tr (c : color) (bh : nat) := 
  if d_down tr then is_redblack tr Red bh.-1 else is_redblack tr c bh.

Lemma balanceL'_Black_deleted_is_redblack l r n c :
  0 < n -> is_deleted_redblack l Black n.-1 -> is_redblack r Black n.-1 ->
  is_deleted_redblack (balanceL' Black l r) c n.
Lemma balanceL'_Red_deleted_is_redblack l r n :
  is_deleted_redblack l Red n -> is_redblack r Red n ->
  is_deleted_redblack (balanceL' Red l r) Black n.
(* similar statements with respect to balanceR' *)
\end{minted}

Regarding leaves, we need special processing in the base cases of
\delete{}, as illustrated in Fig.~\ref{fig:deletion}.
\delete{} might have to ``borrow'' a bit from a sibling of a target
leaf or combine target siblings (possibly after a rotation), to
preserve the size invariants.
Afterwards, \delete{} will recursively rebalance the whole \coqin{dtree}.

Thus we implement \delete{} (as \coqin{ddel}), and prove its
correctness as follows:
\begin{minted}{ssr}
Fixpoint ddel (B : dtree) (i : nat) : deleted_dtree := ...

Lemma ddeleteE B i : wf_dtree' B -> dflatten (ddel B i) = delete (dflatten B) i.
Lemma ddelete_wf (B : dtree) n i :
  is_redblack B Black n -> i < dsize B -> wf_dtree' B -> wf_dtree' (ddel B i).
Lemma ddelete_is_redblack B i n :
  is_redblack B Red n -> exists n', is_redblack (ddel B i) Red n'.
\end{minted}
%where \coqin{ddel} implements \delete{}.
%and \coqin{ddelete} a wrapper that extracts the resulting tree.

These statements are variants of the properties~\preservedata{},
\maintainproperties{} and \returnbalanced{} of Sect.~\ref{sec:insert}.
The proofs are complicated by the huge number of cases,
handled using the proof techniques discussed in the next section.

\section{Using small-scale reflection with inductive data}
\label{sec:proof_techniques}

The small-scale reflection approach is known to be beneficial for
mathematical proofs~\cite{mathcompbook}. However, while \ssreflect{} tactics
are now widely used in the \coq{} community, it is not always clear how
to write proofs of programs using inductive data structures in an
idiomatic style, in particular in presence of deep case analysis.

In the first part of the paper, concerning level-order traversal, the
question is not so acute, as the induction principle we need for LOUDS
is not structural on the shape of trees, but rather on paths,
represented as lists, which are already well supported by the
\ssreflect{} library. Thus the question was the more traditional one of
which definitions to use, so that we can obtain natural lemmas.
This proved to be a time consuming process, which led to gradually
build a library of lemmas, resulting in proofs that match the
intuition, using almost only case analysis and rewriting.

However, the second part, about dynamic bit vectors, uses heavily
structural induction on binary trees, and required developing some
proof techniques to streamline the proofs.

A basic idea of small-scale reflection is to use recursive
Boolean predicates (i.e., recursive computable functions) rather than
inductive propositions. We have already presented two examples:
\coqin{wf_dtree} and \coqin{is_redblack}. Properly designed, they
allow one to prune case analysis by reducing to \coqin{false} on
impossible cases.
On the other hand, they do not decompose naturally in
inductive proofs, which led us first to apply a standard
technique: define a specialized induction principle for trees
satisfying \coqin{wf_dtree} (\coqin{dtree_ind} in Sect.~\ref{sec:basic_queries}).
Using it, the correctness of static queries and
non-structural modification operations (i.e., setting and clearing of
bits) were easy to prove, as the case analysis was trivial.

Properties of \coqin{dinsert}, \coqin{ddel}, and their auxiliary
functions are trickier to prove, as they require complex case analyses
and delicate re-balancing of branches.
Nevertheless, we essentially applied the same principle of
solving goals through direct case analysis.
With this approach, the correctness lemmas (which state that our
operations are semantically correct) were largely automated, consistent with prior research~\cite{nipkow16}. The structural lemmas 
were harder to prove, mainly due to the sheer number of cases
involved and the complexity of invariants.
Our proofs proceed by first applying case analysis to the tree up to the
required depth, and then decomposing all assumptions
to repeatedly rewrite the goal using them until it is solved.
This proof pattern is captured by the following tactic:
\begin{minted}{ssr}
Ltac decompose_rewrite :=
  let H := fresh "H" in case/andP || (move=>H; rewrite ?H ?(eqP H)).
\end{minted}
It is reminiscent of the \coqin{intuition} tactic, a generic tactic
for intuitionistic logic which breaks both hypotheses and goals into
pieces; here we rather rely on rewriting inside Boolean conjunctions to
solve goals piecewise.
For \coqin{dinsert}, this approach instantly finishes most of our proofs,
especially those about red-black tree invariants; the few cases that
require manual treatment being usually handled in one single
\coqin{rewrite}. 
This is true for most auxiliary functions of \coqin{ddel} too,
with one caveat: where \coqin{dinsert} has us generate a dozen cases,
\coqin{ddel} requires hundreds. To cope with this, we
had first to decompose the case analysis in steps, solving most cases
on the way, which means losing some simplicity to speed up proof search.
The proof is still mostly automatic: apply
\coqin{decompose_rewrite}, and throw in relevant lemmas.
When possible, it appears that using \coqin{apply} instead of
\coqin{rewrite} speeds up by a factor of 2 or more, which matters
when the lemma takes more than 1 minute to prove.
We have only 3 such time-consuming case analyses, one for each invariant.
Among the 12 lemmas involved in proving the invariants,
only the inductive proof of well-formedness for
\coqin{ddel} seems to show the limit of this approach, as it
required specific handling for each case of the function definition.

\begin{table}[t]
\centering
  \begin{tabular}{ l | c | c }
    Contents (\coqin{Section} in {\tt dynamic\us redblack.v}) & Lines of code & Lines of proof \\
    \hline
    Definitions (\coqin{btree}, \coqin{dtree}) & 19 & 18\\
    Queries (\coqin{dtree}) & 38 & 58\\
    Insertion (\coqin{insert}, \coqin{dinsert}) & 65 & 208\\
    Set/clear a bit (\coqin{set_clear}) & 25 & 120\\
    Deletion (\coqin{delete}, \coqin{ddelete}) & 98 & 215\\
  \end{tabular}
\caption{Implementation of dynamic bit vectors (see Table~\ref{tab:implementation_overview} for the whole implementation)}
\label{tab:implementation_rb_trees}
\end{table}

\begin{table}[t]
\centering
\begin{tabular}{l|l}
File in~\cite{compact} & Section in this paper \\
\hline
{\tt rank\us select.v} & Sections \ref{sec:counting_rank}, \ref{sec:finding_select}, \ref{sec:theory_rank_select} \\
{\tt pred\us succ.v} & Sect.~\ref{sec:succ} \\
{\tt tree\_traversal.v} & Sections \ref{sec:lo_traversal}, \ref{sec:structural_lo_traversal} \\
{\tt louds.v} & Sections \ref{sec:louds_encoding}, \ref{sec:louds_functions} \\
{\tt dynamic\us redblack.v} & Sect.~\ref{sec:dynamic_vectors} \\
\end{tabular}
\caption{Formalization overview~\cite{compact} (see Table~\ref{tab:implementation_rb_trees} for the details about dynamic bit vectors)}
\label{tab:implementation_overview}
\end{table}

For comparison, Table~\ref{tab:implementation_rb_trees} provides the size of code and proof
required for each \coqin{Section} of our proof script. This does not include lemmas
about the list-based reference implementation.
Note that we count all Boolean
predicates used to model properties as proofs.

The proofs of \vecset{} and \vecclear{}, which we
did not describe here, may seem relatively verbose.
We prove the same properties (a,b,c) as in Sect.~\ref{sec:insert},
but the number of lines hides a disparity between proofs of (a)
correctness and (c) red-blackness, which are almost immediate, as
the structure of the tree is unchanged, and (b) invariants of the
meta-data, for which switching a bit requires to propagate the
difference back to the root, with extra local invariants.

Last, we mention our experience with alternative approaches.
In parallel with our developement using small-scale reflection, we
attempted to formalize dynamic bit vectors using dependent types,
where all invariants are encoded in the type of the data itself.
While this guarantees that we never forget an invariant, difficulties
with the \coqin{Program}~\cite{sozeauthesis} environment led us to write some
functions using tactics~\cite[\coqin{dynamic_dependent_tactic.v}]{compact}.
As written in Sect.~\ref{sec:delete}, this direct connection between
code and proof actually helped us discover some tricky invariants.
However, the resulting code does not lend itself to further analysis,
hence our choice here to stick to a more conventional separation
between code and proof. We did eventually succeed in re-implementing the
dependently-typed version using the \coqin{Program} environment, but at the price of very verbose
definitions~\cite[\coqin{dynamic_dependent_program.v}]{compact}.

Table~\ref{tab:implementation_overview} gives an at-a-glance overview
of our entire Coq development, with a list of files and their
corresponding sections in this paper.

\section{Related work}
\label{sec:related_work}

\coq{} has been used to formalize a constant-time, $o(n)$-space
\rank{} function that was furthermore extracted to efficient OCaml
code~\cite{tanaka2016icfem} and C code~\cite{tanaka2017jip}. This work
focuses on the \rank{} query for static bit arrays while our work
extends the toolset for succinct data structures with more queries
(\select{}, \succ{}, etc.)\ and dynamic structures.

The functions \coqin{level_traversal} and \coqin{lo_traversal_st} of
Sect.~\ref{sec:structural_lo_traversal} match functions given in
squiggle notation in related work by Jones and
Gibbons~\cite{Jones93}. In this work, the \coqin{mzip} function of
Sect.~\ref{sec:structural_lo_traversal} also appears and is called
``long zip with plussle''.
To the best of our knowledge, the function \coqin{lo_traversal_lt} is
original to our work.

Larchey-Wendling and Matthes recently studied the
certification and extraction of breadth-first
traversals~\cite{Larchey2019mpc}. They too define
\coqin{lo_traversal_st}, but then prove it equivalent to a queue based
algorithm, which they extract to efficient OCaml code.
Their goal is orthogonal to ours, as for succinct data
structures what matters is not the efficiency of the traversal, but
the correctness of the parent/child navigation functions, which by
definition require a constant number of queries.

One may use any kind of balanced binary tree to represent dynamic bit
vectors~\cite{navarro2016}. There are many purely-functional balanced
binary search trees, such as AVL trees~\cite{avl} and weight-balanced
trees~\cite{adams93}, but purely functional red-black
trees~\cite{kahrs01,okasaki98} are most widely studied and preferred
by us. As a matter of fact, they have already been formalized in
\coq{}~\cite{appel11,chlipala13,filliatre2004esop},
Agda~\cite{oster11}, and Isabelle~\cite{nipkow16}.

We had to re-implement red-black trees due to the difference of stored
contents.  Above \coq{} formalizations are intended to represent sets,
and maintain the ordering invariant.
Our trees represent vectors, and maintain both that the contents (as
concatenation of the leaves) are unchanged, and that meta-data in
inner nodes is correct (see
Sect.~\ref{sec:representing_dynamic_vectors}).
Still, we found many hints in related work.
For example, in Sect.~\ref{sec:insert} about insertion, the balancing
functions use Okasaki's well-known purely functional balance
algorithm~\cite{okasaki98}, and we formulate our invariants and
propositions similarly to above \coq{} formalizations.

There are now many proofs of programs that use \ssreflect{}, but we
could not find much discussion trying to synthesize the new techniques
put at work.
Sergey et al.\ used \ssreflect{} for teaching
\cite{sergey_pnp,sergey_nanevski_introducing}, observing
benefits for clarity and maintainability, but also giving examples
of custom tactics needed to prove programs.
Gonthier et al.\ \cite{gonthier_ziliani_nanevski_dreyer_2013} have
shown how, in some cases, one can avoid relying on ad hoc tactics
through an advanced technique involving overloading of lemmas.
The techniques we describe in
Sect.~\ref{sec:proof_techniques}, while more rudimentary, are
simple and efficient, yet we have not seen them described elsewhere.

\section{Conclusion}
\label{sec:conclusion}

We reported on an effort to formalize succinct data structures. We
started with a foundational theory of the \rank{} and \select{}
functions for counting and searching bits in immutable arrays.  Using
this theory, we formalized a standard compact representation of trees
(LOUDS) and proved the correctness of its basic operations. Last, we
formalized dynamic bit vectors: an advanced topic in succinct data
structures.

Our work is a first step towards the construction of a formal theory
of succinct data structures. We already overcame several technical
difficulties while dealing with LOUDS trees: it took much care to
find suitable recursive traversals and to sort out the off-by-one
conditions when specifying basic operations.
Similarly, the formalization of dynamic vectors could not be reduced
to the matter of extending conservatively an existing formalization of
balanced trees: we needed to re-implement them to accommodate specific
invariants.

As for future work, we plan to enable code extraction for the
functions we have been verifying, and prove their complexity,
so as to complete previous work~\cite{tanaka2016icfem} and ultimately
achieve a formally verified implementation of succinct data
structures.
%[JG: the following might summarized in a shorter way]
We have already shown that the LOUDS representation of a tree with $n$
nodes uses just $2n$ bits of data. For the LOUDS operations, constant
time complexity is a direct consequence of their being implemented
using a constant number of \rank{} and \select{} operations.  For
dynamic bit vectors, we will first need to properly define a framework
for space and time complexity.

\bibliography{bib}

\begin{thebibliography}{10}

\bibitem{adams93}
Stephen Adams.
\newblock Functional pearls: Efficient sets---a balancing act.
\newblock {\em Journal of Functional Programming}, 3(4):553--561, Oct 1993.

\bibitem{avl}
G.~M. Adel'son-Vel'ski$\breve{\mbox{\i{}}}$ and E.~M. Landis.
\newblock An algorithm for the organization of information.
\newblock {\em Soviet Mathematics--Doklady}, 3(5):1259--1263, September 1962.

\bibitem{compact}
Reynald Affeldt, Jacques Garrigue, Xuanrui Qi, and Kazunari Tanaka.
\newblock A {Coq} formalization of succinct data structures.
\newblock \url{https://github.com/affeldt-aist/succinct}, 2018.

\bibitem{appel11}
Andrew~W. Appel.
\newblock Efficient verified red-black trees.
\newblock Available at
  \url{http://www.cs.princeton.edu/~appel/papers/redblack.pdf} (code included
  in \coq{} standard library, file {\tt MSetRBT.v}, with extra modifications by
  Pierre Letouzey), Sep 2011.

\bibitem{chlipala13}
Adam Chlipala.
\newblock {\em Certified Programming with Dependent Types: A Pragmatic
  Introduction to the Coq Proof Assistant}.
\newblock The MIT Press, 2013.

\bibitem{filliatre2004esop}
Jean{-}Christophe Filli{\^{a}}tre and Pierre Letouzey.
\newblock Functors for proofs and programs.
\newblock In {\em Proceedings of the 13th European Symposium on Programming
  ({ESOP} 2004), Barcelona, Spain, March 29--April 2, 2004}, volume 2986 of
  {\em Lecture Notes in Computer Science}, pages 370--384. Springer, 2004.
\newblock Source code about red-black trees available at
  \url{https://github.com/coq-contribs/fsets/SetRBT.v}.

\bibitem{Gibbons91}
Jeremy Gibbons.
\newblock {\em Algebras for Tree Algorithms}.
\newblock PhD thesis, Programming Research Group, Oxford University, 1991.
\newblock Available as Technical Monograph PRG-94.

\bibitem{ssrman}
Georges Gonthier, Assia Mahboubi, and Enrico Tassi.
\newblock A small scale reflection extension for the {Coq} system.
\newblock Technical report, INRIA, 2008.
\newblock Version 17 (Nov 2016).

\bibitem{gonthier_ziliani_nanevski_dreyer_2013}
Georges Gonthier, Beta Ziliani, Aleksandar Nanevski, and Derek Dreyer.
\newblock How to make ad hoc proof automation less ad hoc.
\newblock {\em Journal of Functional Programming}, 23(4):357–401, 2013.

\bibitem{Jones93}
Geraint Jones and Jeremy Gibbons.
\newblock Linear-time breadth-first tree algorithms: An exercise in the
  arithmetic of folds and zips.
\newblock Technical Report~71, Department of Computer Science, University of
  Auckland, 1993.

\bibitem{kahrs01}
Stefan Kahrs.
\newblock Red-black trees with types.
\newblock {\em Journal of Functional Programming}, 11(4):425--432, Jul 2001.

\bibitem{kudo2011efficient}
Taku Kudo, Toshiyuki Hanaoka, Jun Mukai, Yusuke Tabata, and Hiroyuki Komatsu.
\newblock Efficient dictionary and language model compression for input method
  editors.
\newblock In {\em Proceedings of the Workshop on Advances in Text Input Methods
  (WTIM 2011)}, pages 19--25, 2011.

\bibitem{Larchey2019mpc}
Dominique Larchey-Wendling and Ralph Matthes.
\newblock Certification of breadth-first algorithms by extraction.
\newblock In {\em Proceedings of the 13th International Conference on
  Mathematics of Program Construction}, 2019.

\bibitem{mathcompbook}
Assia Mahboubi and Enrico Tassi.
\newblock {\em Mathematical Components}.
\newblock Available at: \url{https://math-comp.github.io/mcb/}, 2016.
\newblock With contributions by Yves Bertot and Georges Gonthier.

\bibitem{navarro2016}
Gonzalo Navarro.
\newblock {\em Compact Data Structures: A Practical Approach}.
\newblock Cambridge University Press, 2016.

\bibitem{navsad14}
Gonzalo Navarro and Kunihiko Sadakane.
\newblock Fully functional static and dynamic succinct trees.
\newblock {\em ACM Transactions on Algorithms}, 10(3), 2014.

\bibitem{nipkow16}
Tobias Nipkow.
\newblock Automatic functional correctness proofs for functional search trees.
\newblock In {\em Interactive Theorem Proving (ITP 2016)}, volume 9807 of {\em
  Lecture Notes in Computer Science}, pages 307--322. Springer, 2016.

\bibitem{okasaki98}
Chris Okasaki.
\newblock {\em Purely Functional Data Structures}.
\newblock Cambridge University Press, 1998.

\bibitem{oster11}
Julien Oster.
\newblock An {Agda} implementation of deletion in left-leaning red-black trees.
\newblock Available at
  \url{https://www.reinference.net/llrb-delete-julien-oster.pdf}, Mar 2011.

\bibitem{ramanetal01}
Rajeev Raman, Venkatesh Raman, and S.~Srinivasa Rao.
\newblock Succinct dynamic data structures.
\newblock In {\em Proceedings of the 7th International Workshop on Algorithms
  and Data Structures}, pages 426--437. Springer, 2001.

\bibitem{sergey_pnp}
Ilya Sergey.
\newblock Programs and proofs: Mechanizing mathematics with dependent types
  (lecture notes).
\newblock Available at \url{https://ilyasergey.net/pnp/}.

\bibitem{sergey_nanevski_introducing}
Ilya Sergey and Aleksandar Nanevski.
\newblock Introducing functional programmers to interactive theorem proving and
  program verification (teaching experience report).
\newblock Available at \url{https://ilyasergey.net/papers/teaching-ssr.pdf},
  2015.

\bibitem{sozeauthesis}
Matthieu Sozeau.
\newblock {\em Un environnement pour la programmation avec types
  d\'{e}pendants}.
\newblock PhD thesis, Universit\'{e} de Paris-Sud, 2008.

\bibitem{tanaka2016icfem}
Akira Tanaka, Reynald Affeldt, and Jacques Garrigue.
\newblock Formal verification of the rank algorithm for succinct data
  structures.
\newblock In {\em 18th International Conference on Formal Engineering Methods
  (ICFEM 2016), Tokyo, Japan, November 14--18, 2016}, volume 10009 of {\em
  Lecture Notes in Computer Science}, pages 243--260. Springer, Nov 2016.

\bibitem{tanaka2017jip}
Akira Tanaka, Reynald Affeldt, and Jacques Garrigue.
\newblock Safe low-level code generation in {Coq} using monomorphization and
  monadification.
\newblock {\em Journal of Information Processing}, 26:54--72, 2018.

\end{thebibliography}

\end{document}